\documentclass[conference]{IEEEtran}
\IEEEoverridecommandlockouts
\usepackage{cite}
\usepackage{hyperref}
\hypersetup{
	colorlinks=true,
	linkcolor=red,
	filecolor=magenta,      
	urlcolor=cyan,
	citecolor=red
}
\usepackage{enumitem}  

\usepackage{subfig}

\usepackage{physics}
\usepackage{amsmath,amssymb,amsfonts}
\usepackage{graphicx}
\usepackage{textcomp}
\usepackage{varwidth}
\usepackage{xcolor}
\usepackage{multirow}
\usepackage{multicol}

\usepackage{algpseudocode}
\usepackage{algorithm}

\def\BibTeX{{\rm B\kern-.05em{\sc i\kern-.025em b}\kern-.08em
    T\kern-.1667em\lower.7ex\hbox{E}\kern-.125emX}}

\begin{document}

\title{{Network Operations Scheduling \\ for Distributed Quantum Computing}

\author{\IEEEauthorblockN{Nitish Kumar Chandra\IEEEauthorrefmark{1}, Eneet Kaur\IEEEauthorrefmark{2}, Kaushik P. Seshadreesan\IEEEauthorrefmark{1}}
\IEEEauthorblockA{\IEEEauthorrefmark{1} Department of Informatics \& Networked Systems, School of Computing \& Information, \\
University of Pittsburgh, Pittsburgh, PA 15260, USA
\\Emails: nkc16@pitt.edu, kausesh@pitt.edu
}
\IEEEauthorblockA{\IEEEauthorrefmark{2} Cisco Quantum Lab, Los Angeles, CA 90404, USA 
\\Email: ekaur@cisco.com
}
}
}
\maketitle

\begin{abstract}
Realizing distributed architectures for quantum computing is crucial to scaling up computational power. 
A key component of such architectures is a scheduler that coordinates operations over a short-range quantum network required to enable the necessary non-local entangling gates between quantum processing units (QPUs). 
It is desirable to determine schedules of minimum make span, which in the case of networks with constrained resources hinges on their efficient usage. 
Here we compare and contrast two approaches to solving the make span minimization problem, an approach based on the resource constrained project scheduling (RCPSP) framework, and another based on a greedy heuristic algorithm. 
The workflow considered is as follows. Firstly, the computational circuit is partitioned and assigned to different QPUs such that the number of nonlocal entangling gates acting across partitions is minimized while the qubit load is nearly uniform on the individual QPUs, which can be accomplished using, e.g., the METIS solver. 
Secondly, the nonlocal entangling gate requirements with respect to the partitions are identified, and mapped to network operation sequences that deliver the necessary entanglement between the QPUs. Finally, the network operations are scheduled such that the make span is minimized. 
As illustrative examples, we analyze the implementation of a small instance of the Quantum Fourier Transform algorithm over instances of a simple hub and spoke (star) network architecture comprised of a quantum switch as the hub and QPUs as spokes, each with a finite qubit resource budget. 
In one instance, our results show the RCPSP approach outperforming the greedy heuristic. 
In another instance, we find the two performing equally well. Our results thus illustrate the effectiveness of the RCPSP framework, while also underlining the relevance and usefulness of greedy heuristics.


\end{abstract}

\begin{IEEEkeywords}
Distributed Quantum Computing, METIS Algorithm, Resource Constrained Project Scheduling Problem (RCPSP), Greedy Algorithm
\end{IEEEkeywords}

%
\IEEEpeerreviewmaketitle

\section{Introduction}

The field of quantum computing continues to make giant strides in terms of inventions of novel applications with provable speedups~\cite{Dalzell2023-hr}, novel schemes for error correction and fault tolerance~\cite{Bravyi2024-sp}, and maturing hardware technologies with improved coherence times of qubits and gate fidelities~\cite{Temme2017-ih}. 
Quantum computation of any meaningful size in terms of number of qubits can, however, quickly run into large numbers (e.g., tens of thousands to millions in the case of error corrected fault-tolerant computation) that are hard to support with monolithic hardware. 
Techniques such as circuit cutting (wire and gate cutting) and stitching~\cite{circ_cutting, 10.1145/3445814.3446758} attempt to circumvent the issue using a divide and conquer approach, but suffer from steep  
classical computational overheads. 

A more viable approach is to distribute and conquer, i.e., where a large-scale quantum computation is distributed and executed over a collection of finite-sized \emph{quantum processing units} (QPUs) that are interconnected with the help of a quantum network to function as one large computer. 
Distributed quantum computing (DQC)~\cite{CALEFFI2024110672} relies on two key ingredients. 
Firstly, a compiler that i) decomposes the unitary gates in a computational circuit in terms of single and two qubit gates that are native to the hardware, and ii) assigns computational qubits to different QPUs, identifying the entangling gates that are \textit{nonlocal,  teleported} gates, between qubits lying across QPU partitions. 
Secondly, a scheduler that schedules the execution start times of the various gates in the compiled circuit including the nonlocal gates, where the latter involves scheduling the underlying network operations required to enable them.

The quantum circuit compilation-scheduling problem is known to be NP complete even in the case of local computing over monolithic hardware~\cite{Botea2021-gb} that it is at least as hard in DQC if not harder. 
For qubit assignment over QPUs, it is desirable to partition the circuit such that the number of nonlocal entangling gates between qubits at different QPUs is minimized while qubit workloads are nearly evenly distributed across the QPUs. 
This problem maps to a graph partitioning problem, also known to be NP Hard, but can be solved approximately in an efficient manner using, e.g., the METIS solver~\cite{Kloeckner_PyMetis_2022}. Here, two distinct approaches have been explored, namely \emph{qubit cutting} and \emph{gate cutting}~\cite{davis2023distributed}. Whereas the former looks to partition circuits along the lines of qubits so that gate teleportations are minimized, the latter looks for dynamic partitions along the circuits, which along with gate teleportation also allows for qubits to be teleported between modules to minimize subsequent gate teleportations. 

\begin{figure}[hbt!]
    \centering
    \includegraphics[width=0.7\columnwidth]{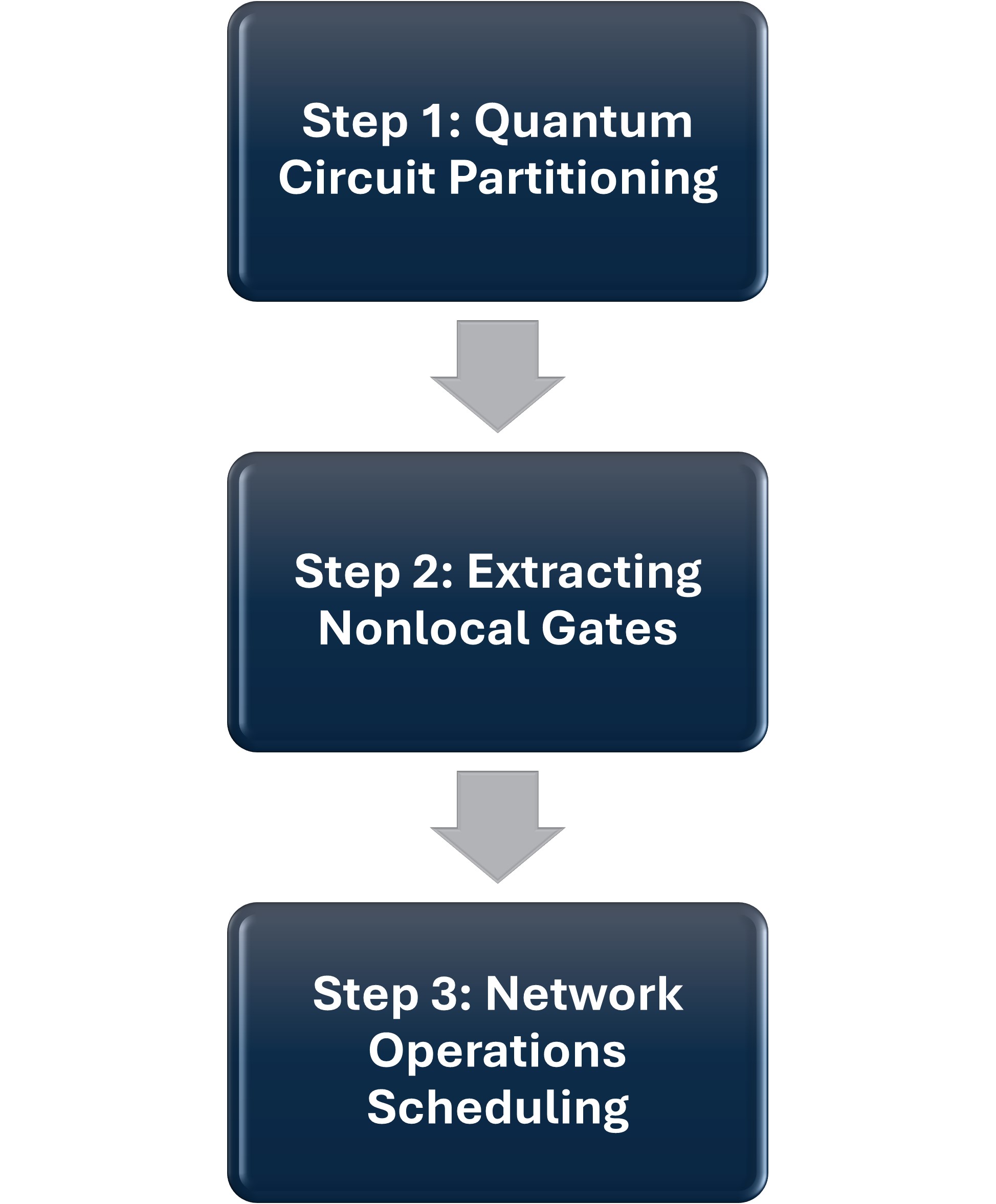}
    \caption{ Workflow for optimizing network operations scheduling in DQC. The input to the process is a compiled quantum circuit that is to be executed over a networked quantum computer, represented in the form of a Gantt chart. The circuit is first partitioned using the METIS solver to minimize the number of nonlocal entangling gates. Secondly, the nonlocal gates are extracted out and mapped to sequences of network operations to generate the necessary entanglement between QPUs. Finally, scheduling approaches based on a Greedy algorithm and on the RCPSP framework are analyzed to find the optimal schedule with the minimum make span.}
    \label{fig:p1}
\end{figure}

The present paper is mainly focused on the final task, namely the task of scheduling the network operations that are necessary to support DQC with a goal of optimizing the make span of the schedule. 
We consider a workflow as described in Fig.~\ref{fig:p1}, which, given a set of assumptions regarding the network architecture, resources, and set of available operations, at the QPUs and network switch nodes, includes the following: i) Quantum circuit partitioning using the METIS solver for distribution over QPUs, ii) Identifying the non-local entangling gates and mapping them to sequences of network operations required to create the necessary entanglement between QPUs that enable the non-local entangling gates, and iii) Optimally scheduling those network operations. We analyze a simple hub and spoke, in other words, a star network topology with a quantum switch acting as the hub and the QPUs as spokes, each with a finite communication and memory qubit resource budget. For the set of network operations consisting of link level entanglement generation, entanglement swapping, and moving entanglement from communication to memory qubits, we study the network operations scheduling associated with executing small circuit instances of the quantum Fourier transform (QFT) algorithm as illustrative examples. To optimize the schedules, we evaluate two methods: one based on the resource-constrained project scheduling problem (RCPSP) framework, and another a greedy heuristic algorithm. Our findings demonstrate that the RCPSP approach is highly effective in generating optimal schedules with minimized make span in resource-limited quantum networks.  They also underline the relevance, usefulness, and efficiency of simple greedy algorithms for the task.

The paper is organized as follows. 
In Sec.~\ref{p1}, we describe quantum circuit representations that are relevant to this work. 
In Sec.~\ref{p3}, we discuss a method for partitioning quantum circuits using the METIS graph partitioning algorithm. In Sec.~\ref{p4}, we describe our two approaches for optimally scheduling quantum operations required to realize the non-local gates. The first is based on the Resource-Constrained Project Scheduling Problem (RCPSP), which focuses on efficiently managing limited quantum resources. The second is a simpler, greedy algorithm that makes locally optimal decisions at each step to minimize execution time. In Sec.~\ref{p5}, we discuss and compare our results from applying these approaches to two simple instances of the star network topology, namely with 2, 4 QPUs. We conclude with a summary in Sec.~\ref{p6}. 



\section{Quantum Circuit Representations}\label{p1}

Two widely used methods for representing quantum circuits are \textit{Directed Acyclic Graphs} (DAGs)~\cite{Treinish2022} and \textit{Gantt charts}~\cite{9810536}, each offering valuable perspectives on the structure and execution of quantum operations. The DAG representation illustrates the dependencies between quantum gates and qubits, where nodes represent either input/output points or operations, and directed edges indicate the flow of qubit data from the output of one gate to the input of another. The DAG representation thus highlights \textit{precedence constraints} that must be satisfied in executing a quantum circuit. Further, it allows key properties like circuit depth to be directly calculated from the graph. Compilers can leverage the DAG representation of a circuit to perform analysis and optimizations for efficient execution on physical hardware, including in distributed settings.

The Gantt chart representation, in contrast, provides a time-based view of quantum circuit execution, with quantum gates represented as tasks spanning specific time intervals. Here, each qubit is treated as a resource, and the chart specifies when each gate operates on a particular qubit. The Gantt chart representation thus clarifies the temporal scheduling of quantum gates, where valid schedules must comply with the precedence constraints specified in the DAG representation, and, additionally, also comply with \textit{resource utilization constraints} that factor in gate latency and qubit availability, and \textit{timing overlap constraints}, namely that no two operations on the same qubit can overlap in time. By clearly illustrating the sequential and parallel execution of operations, the Gantt chart representation thus helps in representing efficient schedules for quantum circuits, especially in systems requiring nonlocal operations or dealing with resource limitations.

\begin{figure}[hbt!]
    \centering
    \includegraphics[width=\columnwidth]{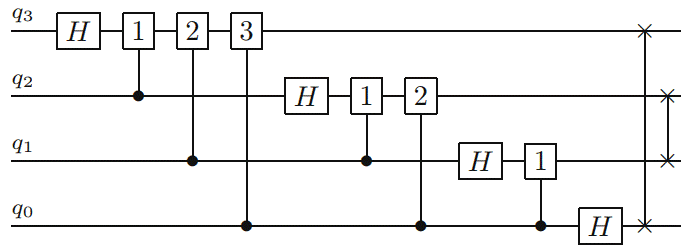}
    \caption{Quantum circuit diagram for a 4-qubit Quantum Fourier Transform (QFT). The circuit begins with Hadamard gates, followed by controlled phase rotations with decreasing angles, and concludes with optional SWAP gates to reverse the qubit order. This circuit transforms the input quantum state from the computational basis to the Fourier basis, a key operation in various quantum algorithms.}
    \label{fig:plot2}

\end{figure}

\begin{figure}[hbt!]
    \centering
    \includegraphics[width=4.5cm]{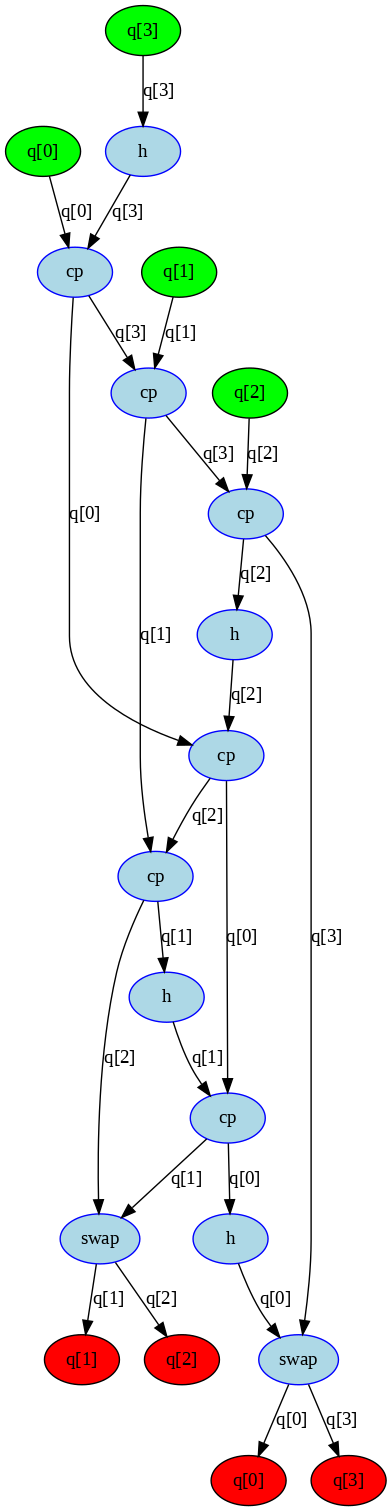}
   \caption{Directed Acyclic Graph (DAG) representation of a 4-qubit Quantum Fourier Transform (QFT) circuit. The nodes represent quantum gates applied during the QFT process, including Hadamard (h) gates, Controlled Phase (cp) gates with decreasing angles, and Swap gates to reverse the order of qubits.}

    \label{fig:plot3}
\end{figure}

Consider the 4-qubit Quantum Fourier Transform (QFT) circuit shown in Fig.~\ref{fig:plot2}. 
Here, $H$ represents the Hadamard gate, and the controlled-phase rotation gates (labeled 1, 2, 3) 
represent rotations by angles \(\pi/2\), \(\pi/4\), and \(\pi/8\) about the $Z$ axis, respectively.  
The DAG representation of the circuit is as shown in Fig.~\ref{fig:plot3}. It is organized to flow from top to bottom and left to right, mirroring the sequential progression of the circuit execution. 
The controlled-phase rotations, denoted by \textit{cp}, are arranged in a cascading manner, with each gate depending on the completion of previous operations. The SWAP gates at the end of the DAG allow for the reordering of qubits if necessary.


\begin{figure*}[ht]
\centering
\includegraphics[width=  16 cm]
{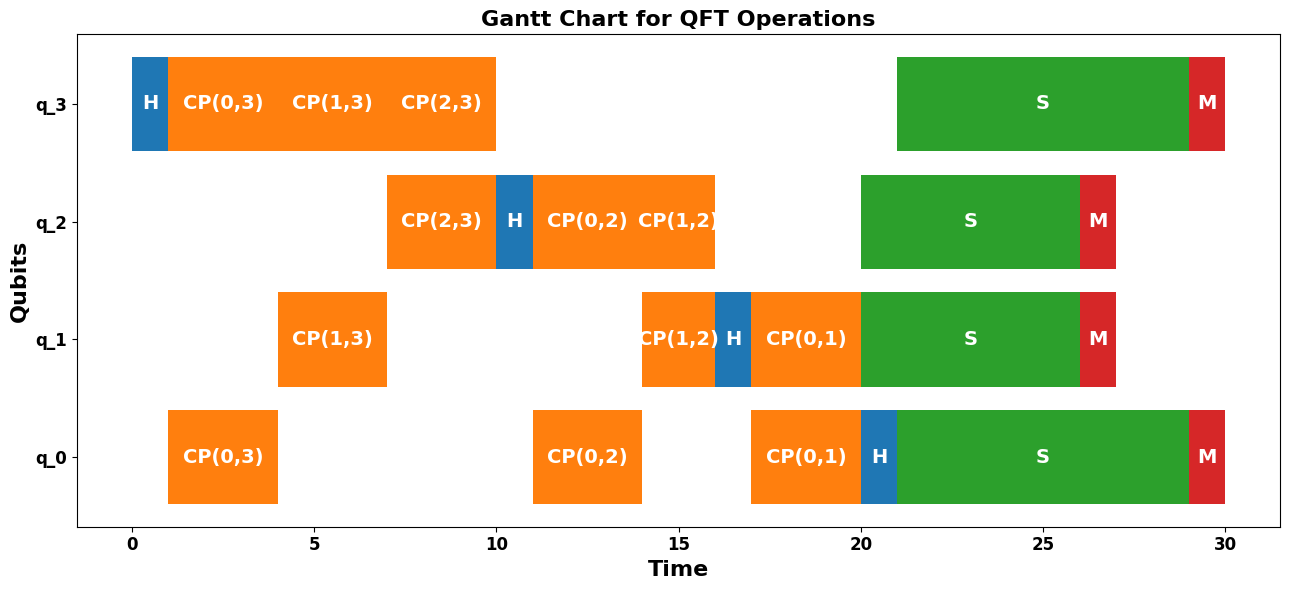}
   \caption{Gantt chart representation of a 4-qubit Quantum Fourier Transform (QFT) circuit, illustrating the scheduled execution of Hadamard (H), controlled-phase (CP) which is in the format CP (Control Qubit, Target Qubit), and SWAP (S) gates. The chart ensures no overlap on the same qubit and synchronizes gate durations to optimize the overall execution time.}
   \label{gantt}
\end{figure*}




A Gantt chart representation of the same circuit is as shown in Fig.~\ref{gantt}. Here, each controlled-phase gate is denoted as \textit{CP(control\_qubit, target\_qubit)}, where the operation depends on the state of the control qubit, affecting the target qubit with a phase rotation. The SWAP gates, represented by \textit{S} that can be implemented as three consecutive CNOT gates, enables the exchange of qubit states. Note that in the schedule depicted in this example Gantt chart, the Hadamard (H) gates, controlled-phase (CP) gates, and SWAP (S) gates are scheduled to ensure that no two operations on the same qubit overlap in time.
Here, single-qubit gates, such as the Hadamard gates, are assumed to take less time than two-qubit gates, like the controlled-phase and SWAP gates. To prevent any overlap, the start time for each gate is carefully calculated by ensuring that it begins only after the completion of all prior operations on the involved qubits. This is achieved by determining the maximum end time of the previous operations on the qubits affected by the new gate. This approach ensures synchronization across the qubits, optimizing the circuit's overall execution time.

\section{Distributing Quantum Circuits for Networked Realization}
\label{p3}
Quantum circuits are inherently complex, often involving many qubits and operations that require careful management of quantum resources. The challenge in DQC is to distribute computational qubits among several QPUs in such a way that the number of non-local operations, i.e., those requiring entanglement between different QPUs is minimized. It can be accomplished using the METIS graph partitioning algorithm~\cite{Kloeckner_PyMetis_2022}. The methodology and application of the METIS algorithm is summarized below.

\subsubsection{Mapping Quantum Circuits to Graphs}

\begin{figure}[hbt!]
    \centering
    \includegraphics[width=8cm]{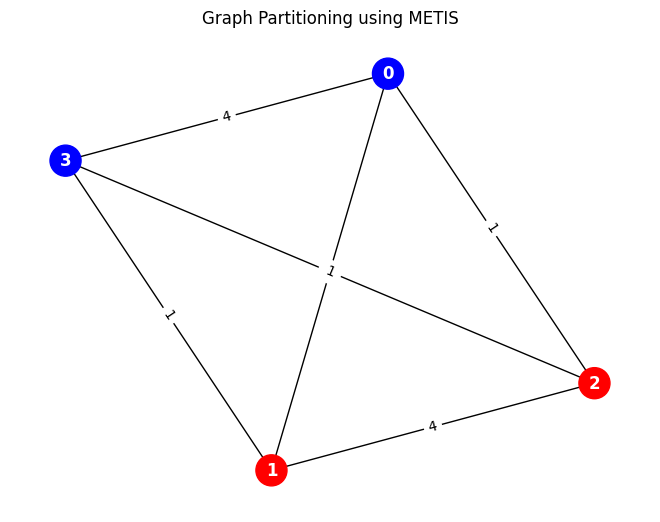}
    \caption{METIS algorithm implemented to distribute 4-qubit QFT circuit in 2 QPUs. The algorithm divides it such that the edges with maximum weight are in one QPU, so qubits (0 \& 3) are in one QPU and qubits (1 \& 2) are in the other QPU in order to minimize non-local controlled gates. }
    \label{fig:plot1}
\end{figure}

For the approach called qubit partitioning~\cite{davis2023distributed}, we begin by representing the quantum circuit as a graph, where each qubit is mapped to a node. The edges between nodes correspond to entangling gates, with the weight of each edge representing the number of entangling gates between the connected qubits. This graph representation captures the essential structure of the quantum circuit and serves as the input to the METIS algorithm.

\subsubsection{METIS Algorithm Overview}

The METIS algorithm is a robust method for partitioning graphs into multiple subgraphs with balanced sizes while minimizing the edge cut, which represents the number of non-local gates. It achieves this by first coarsening the graph—merging nodes to reduce its size—before applying its partitioning algorithm to the simplified version. The final partition is then refined through an uncoarsening process, ensuring that the distribution of qubits across QPUs is optimized~\cite{karypis1997metis}. The original time complexity of METIS algorithm is $O(n + m + k \log(k))$, where $n$ represents the number of vertices, $m$ the number of edges, and $k$ the number of balanced partitions. However, various modified implementations demonstrate improved time complexity compared to the standard METIS algorithm \cite{7752208}. The algorithm's objective is twofold: to minimize the communication overhead between QPUs and to maintain balanced partition sizes.

\subsubsection{Application to Quantum Circuits}

By applying the METIS algorithm to the graph representation of a quantum circuit, we achieve an optimal division of qubits among different QPUs. For example, when applied to distribute the 4-qubit QFT circuit across two QPUs, the optimal partitioning that groups qubits belonging to edges with the highest weights in the same QPU, and is given by $\{$qubit 0, qubit 3$\}$ in one QPU, and $\{$qubit 1, qubit 2$\}$ in the other (See Fig.~\ref{fig:plot1}).

\section{Network Operations Scheduling}\label{p4}

In a networked quantum computer comprised of QPUs and quantum switches~\cite{10313694}, scheduling of entanglement-generating operations is vital for implementing the non-local entangling gates as part of DQC. The underlying quantum networks are often constrained by limited quantum resources at the nodes for networking, i.e., consist of finite number of communication channels and qubits, and memory qubits for storage. Effective utilization of the resources is key to delivering the entanglement copies necessary for the execution of nonlocal gates between QPUs in a timely manner. The trait of ``timeliness" here can be considered at two different levels of abstractions, i) delivering entanglement between different pairs of QPUs in the same order as that of the requirements for nonlocal gates in executing a DQC circuit, and ii) delivering entanglement between QPUs at the precise time of requirement for the execution of each nonlocal gate. Given that the latter may not always be possible to meet under network resource constraints, we stick to the first level of abstraction. Our goal is thus to optimize the make span of the full sequence of network operations required to deliver the necessary entanglement copies for all the nonlocal entangling gates in DQC execution of a quantum circuit. 
In this regard, we explore the use of the Resource-Constrained Project Scheduling Problem (RCPSP) framework alongside a greedy heuristic algorithm.

\subsection{ Resource-Constrained Project Scheduling Problem}

We present a concise overview of the Resource-Constrained Project Scheduling Problem (RCPSP)~\cite{BRUCKER19993,artigues2013resource} and its mathematical formulation to address the challenge of scheduling operations for entanglement generation in quantum networks. RCPSP is a framework that schedules a set of activities, each requiring specific resources, while adhering to constraints and precedence relationships. By mapping our quantum network scheduling problem to RCPSP, we define quantum operations (entanglement generation, entanglement swapping, movement of entanglement pair from communication to memory ions) as activities and the limited quantum resources (communication and memory ions) as the constrained resources. The mathematical formulation ensures that operations are scheduled without exceeding resource capacities and that dependencies are respected. The goal is to develop a schedule that respects resource capacities and operational dependencies, enabling the timely execution of quantum operations while minimizing the overall time required, also known as the make span.



For an activity $j$, let $S_j$ denote the start time and $d_j$ denote the duration. Each activity $j$ requires a specific amount of resource $r$, denoted by $R_{j,r}$, and the total available amount of each resource type is denoted by $B_r$. The primary objective is to minimize the make span, $C_{\text{max}}$, which is defined as the maximum finish time across all activities. Mathematically, the objective function is expressed as
\begin{align}
\min \ C_{\text{max}} = \max(S_j + d_j),
\end{align}
subject to resource and precedence constraints. The resource constraints ensure that the total resource usage at any time $t$ does not exceed the available resource capacity $B_r$, and precedence constraints guarantee that dependent operations are executed in the correct sequence, where one activity can only begin after its predecessor is completed.

To ensure that a schedule is both feasible and optimal, each activity must be assigned a unique start time, subject to the availability of the necessary resources. The resource usage at any time step $t$ is constrained by:
\begin{align}
\sum_{j \in A(t)} R_{j,r} \leq B_r \quad \forall r, t
\end{align}
where $A(t)$ represents the set of activities in progress at the time $t$. Additionally, precedence relations between activities are maintained by enforcing the constraint $S_k \geq S_j + d_j$ for all dependent activities $(j,k)$. By using RCPSP optimization techniques, such as Mixed-Integer Programming (MIP) or Constraint Programming (CP), an optimal schedule that minimizes the make span can be computed that ensures efficient resource allocation while adhering to the dependencies between quantum operations. In this work, we use PuLP, a linear and mixed integer programming modeler for Python~\cite{pulp2023}, to solve for an optimal schedule.

\subsection{Greedy Algorithm for Quantum Network Scheduling}

For comparison, we also analyze a greedy algorithm~\cite{Akcay2007-rg,cormen2022introduction} for efficient scheduling of the quantum network operations for DQC. The algorithm is designed to systematically allocate time slots to various activities while adhering to the precedence constraints and resource availability. 
The greedy approach operates by iteratively scheduling activities, starting from time zero and advancing through the time horizon until all operations are scheduled, or the available time is exhausted. At each time step, the algorithm assesses which activities are eligible for scheduling based on their readiness and the availability of quantum resources, such as communication and memory qubits. 
Once an activity meets the necessary conditions, it is scheduled to commence at the current time step, with its duration determined by its predefined length. The required resources for the activity are reserved for the entirety of its duration, preventing other activities from accessing these resources simultaneously. The algorithm continues to adjust the availability of resources and the start times of activities until the scheduling process is completed, or the time horizon is reached. This approach aims to optimize scheduling efficiency, enabling the successful execution of non-local quantum gates. 
The pseudocode for the algorithm is described below as Algorithm 1.

\begin{algorithm}
\caption{Greedy Scheduling Algorithm}
\begin{algorithmic}[1]
\State \textbf{Input:} Number of activities $n$, time horizon $T$, durations $d$, precedence constraints, resource availability $B$, resource requirements $r$
\State \textbf{Output:} Start times of activities

\State Initialize \textit{start\_times} as an array of $-1$ of size $n$
\State Initialize \textit{resource\_availability} as a $|B| \times T$ matrix with each entry set to the available resource amount
\State Initialize \textit{completed\_activities} as an empty set

\For{time $\gets 0$ \textbf{to} $T-1$}
    \State \textit{ready\_activities} $\gets$ Find all activities where all predecessors are completed
    \For{each activity in \textit{ready\_activities}}
        \If{activity is not scheduled}
            \State Determine \textit{earliest\_start\_time} based on precedence constraints
            \If{\textit{earliest\_start\_time} + duration of activity $\leq T$}
                \State Check if resources are available from \textit{earliest\_start\_time} to \textit{earliest\_start\_time} + duration
                \If{resources are available}
                    \State Schedule the activity at \textit{earliest\_start\_time}
                    \State Update \textit{resource\_availability}
                    \State Add activity to \textit{completed\_activities}
                \EndIf
            \EndIf
        \EndIf
    \EndFor
\EndFor

\State \Return \textit{start\_times}
\label{algorithm}
\end{algorithmic}
\end{algorithm}

The greedy scheduling algorithm aims to make locally optimal decisions at each step to approximate a globally optimal solution~\cite{cormen2022introduction,RYBIN2024100824}. 
While the best possible schedule is not always guaranteed, it typically yields a practical, near-optimal solution, especially when the constraints and dependencies of the problem are well-suited to its criteria. Its strength lies in a straightforward implementation and capacity to swiftly adapt to changing conditions in resource availability and precedence constraints, making it a useful method for addressing complex and dynamic scheduling problems.


\section{Results}\label{p5}

\begin{figure*}
    \centering
    \subfloat[\centering Two QPUs and a Quantum Switch  ]{{\includegraphics[width=8cm]{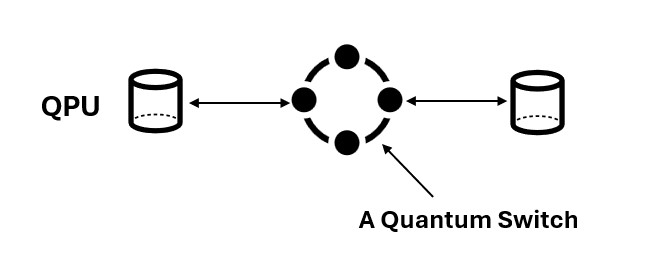} }}%
    \qquad
    \subfloat[\centering Four QPUs and a Quantum Switch]{{\includegraphics[width=8cm]{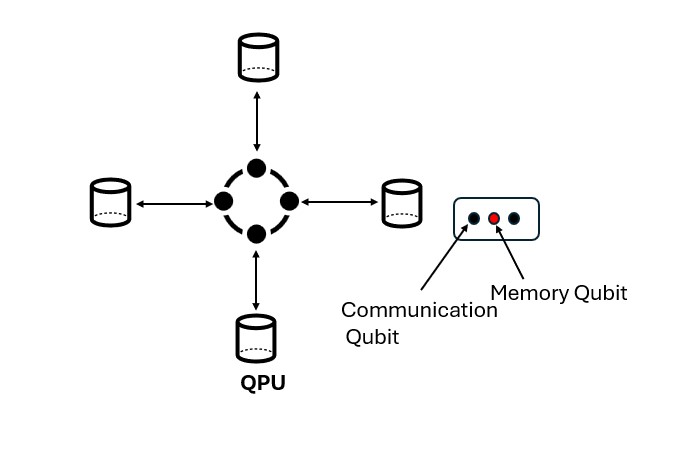} }}%
     \caption{A star topology network for Distributed Quantum Computing (DQC) consists of a central switch connected to multiple Quantum Processing Units (QPUs). In configuration (a), the switch is linked to two QPUs via entangled pairs, while configuration (b) shows a switch linked to four QPUs. Both the switch and the QPUs contain communication qubits (depicted in red), and the QPUs alone contain memory qubits (depicted in black).}
     \label{fig:topology}

\end{figure*}

\begin{figure*}[hbt!]
    \centering
    \includegraphics[width= 14 cm]{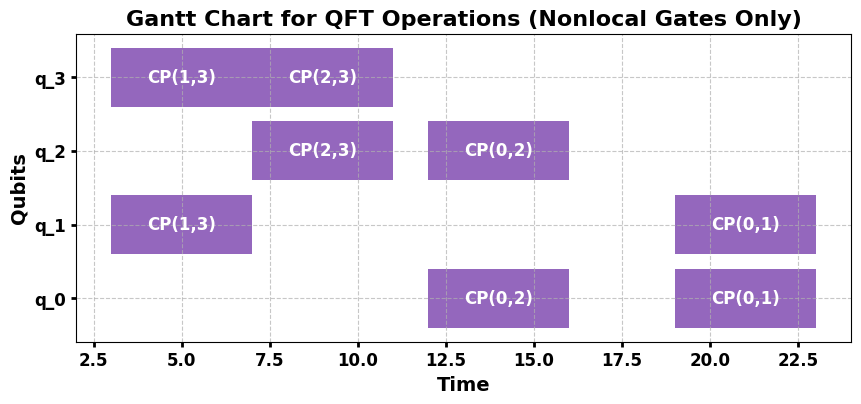}
   \caption{Gantt chart for scheduling non-local operations in a 4-qubit Quantum Fourier Transform (QFT) circuit.}

    \label{fig:plot6}
\end{figure*}

We now focus on a networked quantum computer of star topology as depicted in Fig.~\ref{fig:topology} (depicts two cases: one with a quantum switch and two QPUs, and the other with a quantum switch and four QPUs.). 
The communication qubits depicted in red are employed at both the switch and the QPUs for quantum communication and entanglement generation, while the memory qubits depicted in black are employed at the QPUs alone for storing entangled pairs.
Our goal is to optimally schedule the network operations required to generate all the necessary entanglement pairs between QPUs to enable the nonlocal gates in DQC implementation of the 4-qubit QFT circuits. The Gantt chart in Fig.~\ref{fig:plot6} shows the non-local gates in the 4-qubit QFT circuit for the bi-partition of the qubits between two QPUs, namely $QPU_{1}$ containing qubits 0 and 3, while $QPU_{2}$ containing qubits 1 and 2. Recall that this allocation was determined using the METIS algorithm to minimize the number of non-local gates required.

\begin{figure*}
    \centering
    \subfloat[\centering Schedule obtained by RCPSP algorithm  ]{{\includegraphics[width=8.5cm]{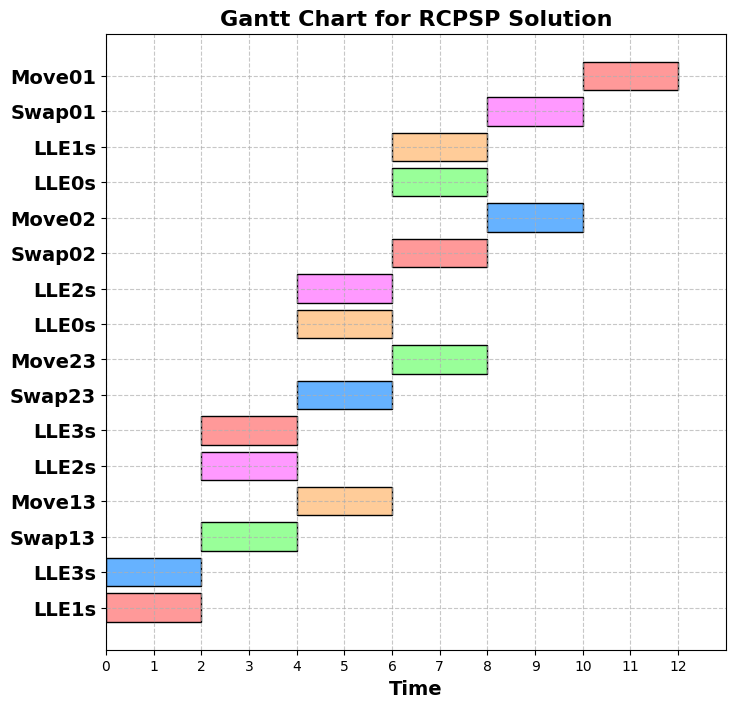} }}%
    \qquad
    \subfloat[\centering Schedule obtained by a Greedy Algorithm ]{{\includegraphics[width=8.5cm]{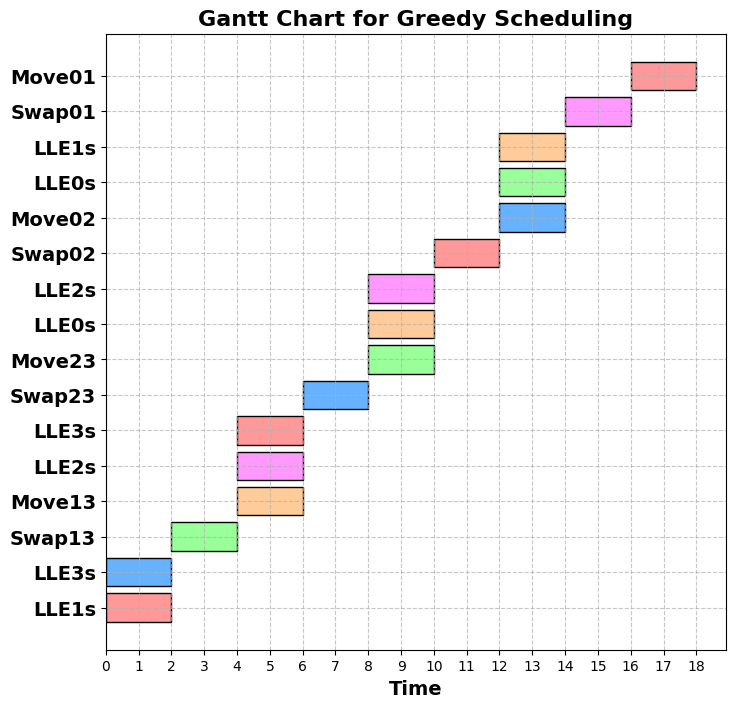} }}%
\caption{Plots for case of two QPUs and a Quantum Switch}
     \label{example1}

\end{figure*}

\begin{figure*}

    \centering
    \subfloat[\centering Schedule obtained by RCPSP algorithm ]{{\includegraphics[width=8.2cm]{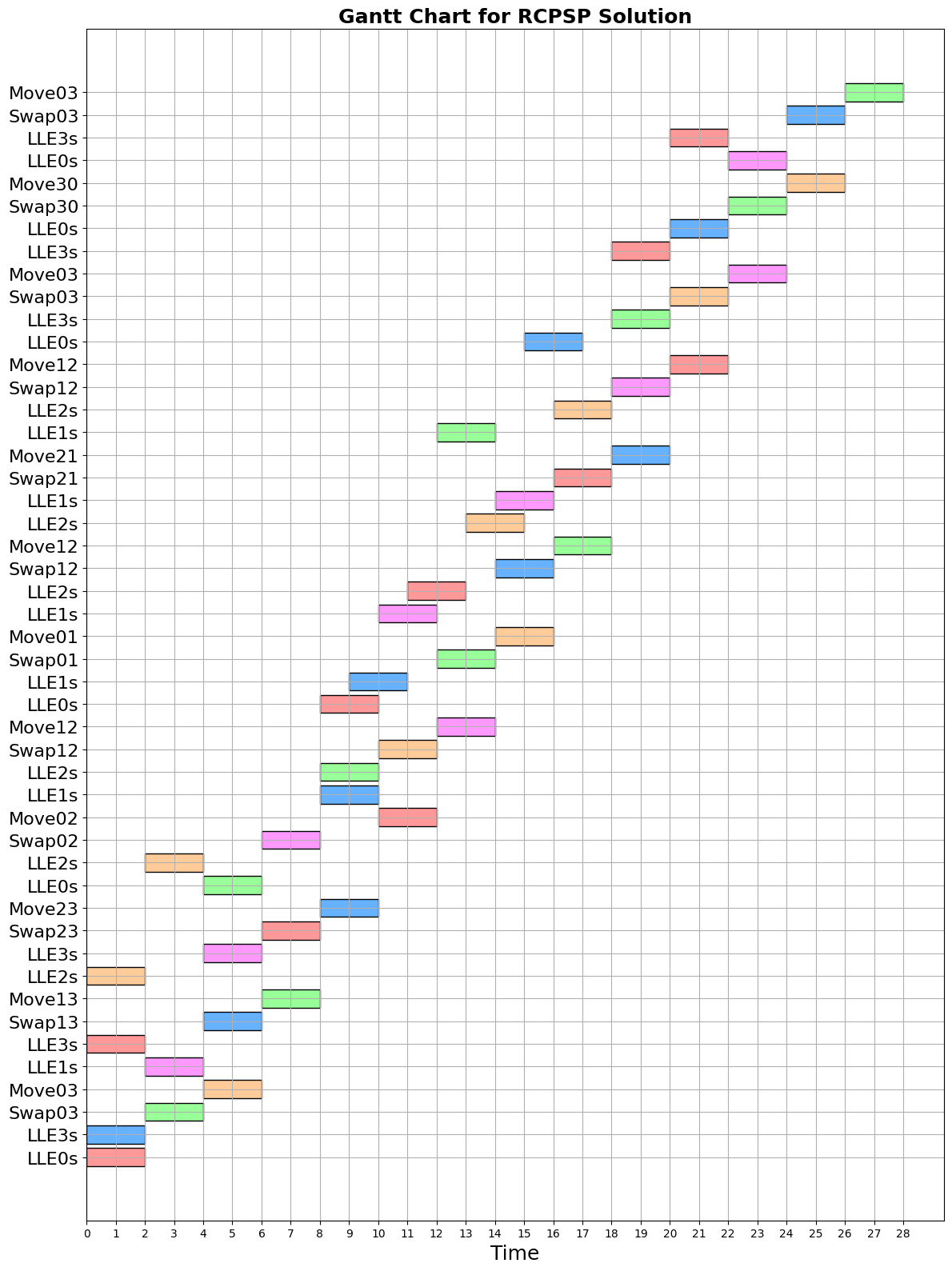} }}%
    \qquad
    \subfloat[\centering Schedule obtained by a Greedy Algorithm ]{{\includegraphics[width=8.5cm]{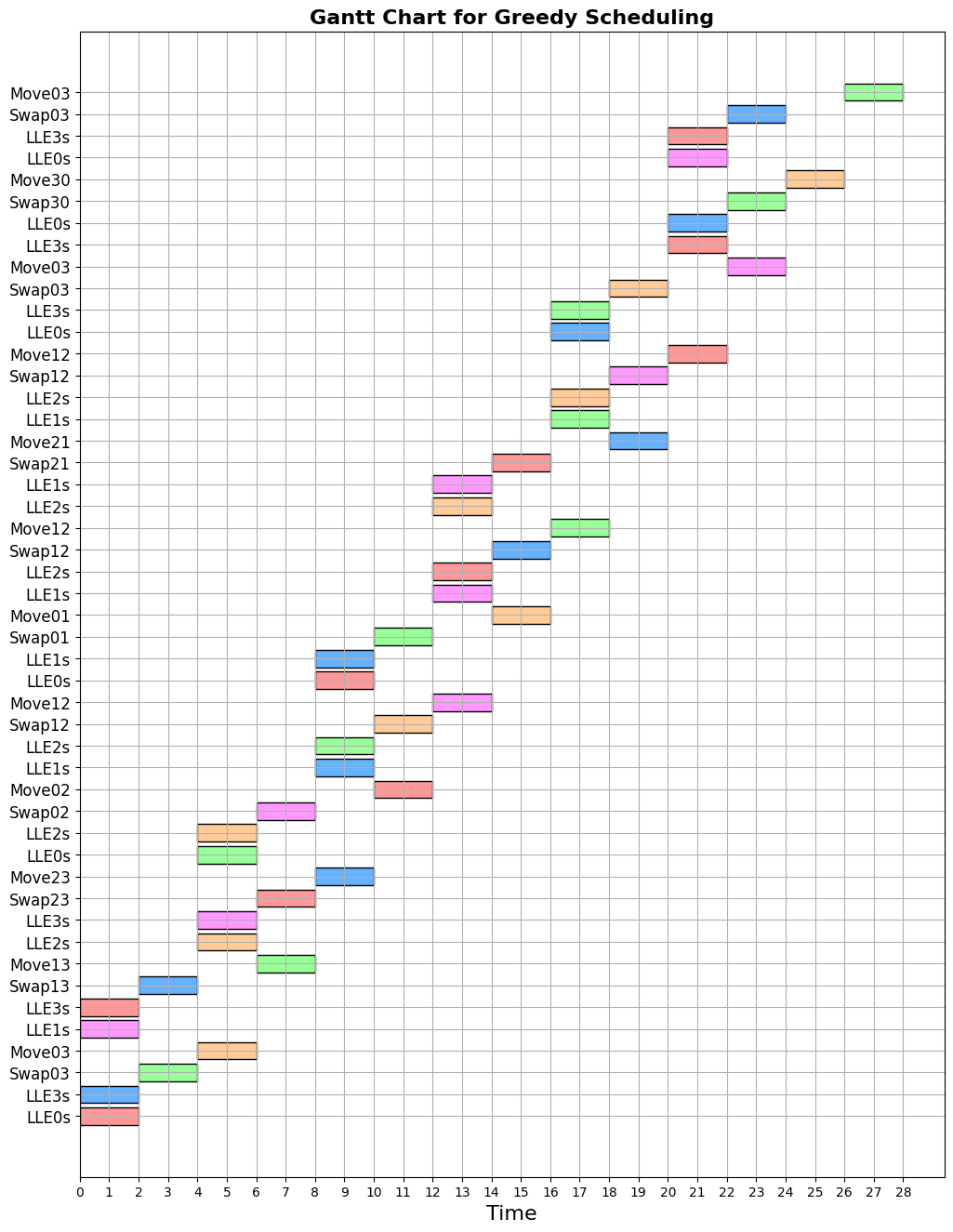} }}%
    \caption{Plots for the case of Four QPUs and a Quantum Switch}%
    \label{fig:example2}
\end{figure*}




In-order to generate end to end entanglement links between QPUs, we consider the following quantum operations,

\begin{enumerate}
    \item \textbf{Link Level Entanglement (LLE):} 
    For a pair of QPUs $i, j$, link-level entanglements are generated between a qubit at each of the QPUs (\( q_i, q_j \)) and a qubit each at the quantum switch. The operation that creates an entangled pair between QPU $i$ and the quantum switch is, e.g., denoted as \( \text{LLE}_{i\text{s}} \).

    \item \textbf{Swap Operation (Swap):}
    Next, a swap operation, \( \text{Swap}_{ij} \), is performed at the quantum switch. This operation transfers the entanglement from the quantum switch to the target qubits \( q_i \) in $QPU_{1}$ and \( q_j \) in $QPU_{2}$, creating an entanglement pair between \( q_i \) and \( q_j \).

    \item \textbf{Move Operation (Move):}
    Finally, a move operation, \( \text{Move}_{ij} \), is executed to transfer the entangled qubits from the quantum switch to their respective memory qubits in $QPU_{1}$ and $QPU_{2}$. This operation ensures that the entangled qubits have enhanced decoherence time within their respective QPUs. 
\end{enumerate}

The details of these operations are similar to those described in Refs.~\cite{Santra_2019,skrzypczyk2021architecturemeetingqualityofservicerequirements}. These sequence of operations effectively establishes a quantum entanglement pair across two QPUs, enabling the execution of non-local quantum gates between distant qubits. 
There are two layers of precedence requirements when scheduling quantum operations. First, link-level entanglements must be generated before performing entanglement swapping at the quantum switch, and move operations at QPUs can only occur after the entanglement swapping is completed. This ensures that quantum entanglement is established in order to operate non-local quantum gates. Second, the move operations must follow a specific sequence, as indicated by the Gantt chart for non-local gates, to avoid conflicts and ensure the correct execution order is followed.



We analyze two exemplary scenarios using both the RCPSP solver and a greedy algorithm, namely star networks with 2 and 4 QPUs, respectively. The greedy algorithm selects tasks based on straightforward criteria like earliest availability, while the Resource-Constrained Project Scheduling Problem (RCPSP) approach accounts for both resource constraints and task dependencies to optimize the schedule. The RCPSP method is designed to manage communication and memory qubits more effectively, ensuring better utilization of available resources.

\subsection{Two QPUs  and a Quantum Switch}

In this case, we consider a quantum switch and two QPUs and schedule quantum operations for nonlocal gates in order to implement four qubit QFT circuit. In this quantum scheduling scenario, the activities represent tasks to be performed across a network of quantum processing units (QPUs) and a quantum switch. Each QPU has two communication qubits, used for generating entanglement pairs with the switch, and two memory qubits, which store these entangled pairs. In contrast, the switches have only two communication qubits available for generating entanglement between different QPUs. Each operation—whether it is a LLE, Swap, or Move—requires two discrete time steps to complete. 

We need to implement four non-local quantum gates, each requiring the generation of an entanglement pair. For each pair, four quantum operations are necessary: two link-level entanglement operations, one quantum swap operation, and one move operation. Since we need to establish four such entangled pairs between the QPUs, this results in a total of 16 quantum
 operations to complete the required four-qubit QFT circuit distributed across two QPUs.

We observe that the RCPSP approach outperforms the Greedy algorithm (See Fig.~\ref{example1}) by producing a more efficient schedule, as demonstrated by the comparison plot. By carefully optimizing the allocation of qubits over time, RCPSP makes more effective use of the limited resources at the QPUs and switches. However, while RCPSP delivers a better solution in this scenario, it does not always guarantee superior performance in every case. In less complex scenarios, where resource constraints are less stringent or task interdependencies are simpler, the Greedy algorithm may suffice and may even be preferable due to its lower computational overhead. 

\subsection{Four QPUs and a Quantum Switch}

In this case, we consider a network consisting of four Quantum Processing Units (QPUs) and a quantum switch. Unlike the two QPU bi-partition case (Fig.~\ref{fig:plot6}), all entangling gates (in Fig.~\ref{gantt}) here can be trivially identified to be nonlocal, which significantly impacts the scheduling process. Specifically, we need to schedule 12 nonlocal gates, resulting in a total of 48 quantum operations. These operations include 2 Link level Entanglements (LLEs) to connect each QPU to the quantum switch, a quantum swapping operation at the switch, and a move operation to transfer the entangled pair from communication qubits to memory qubits. 

In this configuration, each QPU has four communication qubits for establishing entanglement pairs with the switch, and four memory qubits for storing these pairs. On the other hand, the switches have only four communication qubits available, which are dedicated to creating entanglement between different QPUs. Every operation, including Link-Level Entanglement (LLE), Swap, or Move, takes two discrete time steps to execute.

Our analysis shows that the RCPSP approach achieves the same make span  as the Greedy algorithm for this case, as depicted in Fig.~\ref{fig:example2}. Although both methods reach the same final make span, their scheduling strategies differ significantly. The RCPSP method employs a holistic strategy, integrating resource constraints and task dependencies to develop an optimal sequence of operations. It aims to balance these elements to ensure overall efficiency throughout the schedule. In contrast, the Greedy algorithm operates by focusing on immediate, short-term gains, prioritizing tasks based on localized criteria rather than the overall impact on the schedule. The fact that both methods result in the same make span highlights that, despite the Greedy algorithm's focus on individual task optimization, it can achieve comparable results to the more comprehensive RCPSP approach in certain scenarios.

\section{Summary, Conclusions and Outlook}\label{p6}
In summary, to reduce execution time of DQC over a resource-limited networked quantum computer, we examined two strategies for scheduling the necessary network operations, namely one based on a greedy algorithm and another based on the well established Resource-Constrained Project Scheduling Problem (RCPSP) framework. Our approach involved mapping quantum circuits onto graphs, where qubits were represented as nodes and entangling gates as edges. We then divided the circuits across multiple QPUs, striving to evenly distribute qubit load while minimizing the non-local entangling gates needed between partitions, using the METIS solver. These entangling gates identified as non-local gates were subsequently mapped to sequences of network operations to establish the required entanglement links between QPUs. For the four qubit Quantum Fourier Transform (QFT) implemented on a star network topology of QPUs with a quantum switch in the middle, where the QPUs and the switch have limited qubit resources, we employed both the RCPSP framework and a greedy algorithm-based approach to determine optimal network operations schedules.

To conclude, the comparison between the RCPSP approach and the Greedy algorithm highlights important trade-offs in quantum operation scheduling. For a quantum switch with two QPUs, the RCPSP approach clearly outperforms the Greedy algorithm, delivering a more efficient schedule by optimizing qubit allocation and making better use of limited resources. However, this advantage does not extend to all cases. In the scenario involving a quantum switch with four QPUs, both methods achieve the same make span despite their differing approaches. While RCPSP offers a holistic strategy by considering resource constraints and task dependencies, the Greedy algorithm’s focus on immediate gains can still produce comparable results in some scenarios. These findings suggest that the choice between the two approaches should be based on the complexity of the system and available resources.

Both RCPSP and Greedy algorithms have their strengths and limitations, making them suitable for different types of scheduling challenges. RCPSP excels in complex, resource-constrained scenarios, providing more optimized schedules by thoroughly exploring multiple possibilities. However, this comes at the cost of higher computational complexity and longer execution times, which can hinder scalability for larger systems. On the other hand, the Greedy algorithm provides a faster, simpler alternative that is effective for tasks where the computational complexity of RCPSP might be prohibitive. While it often sacrifices optimality due to its focus on immediate, localized decisions, it can still yield near-optimal solutions for practical purposes, making it a viable option when quick results are needed.

In an extended version of this work, to further establish our findings, we intend to evaluate the two scheduling approaches discussed in this paper for the execution of real quantum circuits from a benchmark set~\cite{revkit}. Another direction of interest is to explore the co-optimization of circuit compilation and network operations scheduling to minimize the make span of executing quantum circuits over networked QPUs, where it is not necessary to minimize the absolute number of nonlocal gates in the partitioning of the circuit, but rather only the depth of nonlocal gate layers in the circuit compilation which has a strong dependence on the size of network resources. Furthermore, it is essential, and also interesting, to consider security aspects of DQC over networked quantum computers such as QPU authentication that can confirm the identity of QPUs, and error correction mechanisms that can mitigate noise and ensuing entanglement infidelities, achieving high levels of reliability of quantum communication, which we leave for future work. 

\bibliographystyle{IEEEtran}
\bibliography{references.bib}

\end{document}